\documentclass[aps,prb,twocolumn,superscriptaddress,showpacs]{revtex4}
\usepackage[ansinew]{inputenc}
\usepackage[pdftex]{graphicx}
\usepackage{amsmath}
\begin{document}
\setlength\topmargin{-0.5in}
\setlength\oddsidemargin{-0.37in}

\title{The quantum scattering time and its implications on scattering sources in graphene}

\author{X. Hong}
\affiliation{Department of Physics, The Pennsylvania State
University, University Park, PA 16802}

\author{K. Zou}
\affiliation{Department of Physics, The Pennsylvania State
University, University Park, PA 16802}

\author{J. Zhu}
\affiliation{Department of Physics, The Pennsylvania State
University, University Park, PA 16802}

\date{\today}

\begin{abstract}
We determine the quantum scattering time $\tau_q$ in six graphene
samples with mobility of 4,400 $< \mu < $ 17,000 cm$^2$/Vs over a
wide range of carrier density (1.2 $< n <$ 6x10$^{12}$/cm$^2$).
$\tau_q$ derived from Shubnikov-de Haas oscillations ranges
$\sim$25-74 fs, corresponding to a single-particle level broadening
of 4.5-13 meV. The ratio of the transport to quantum scattering time
$\tau_t/\tau_q$ spans 1.5-5.1 in these samples, which can be
quantitatively understood combining scattering from short-ranged
centers and charged impurities located within 2 nm of the graphene
sheet. Our results suggest that charges residing on the SiO$_2$
surface play a dominant role in limiting carrier mobility in current
samples.

\end{abstract}
 \maketitle

Understanding and eliminating extrinsic scattering sources in
graphene is critical to the advancement of its fundamental study and
technological applications. Despite many theoretical and
experimental investigations into possible candidates, including
charged impurities (CI), adsorbates, substrate corrugations, and
ripples, contradictory observations remain and a clear picture has
yet to emerge.\cite{AndoGraphene,Nomura2007,AdamSelf,WiesendangerResonance,GuineaMidgap,KatsnelsonRipple,KatsnelsonCluster,TanMobility,SchedinGasMolecule,BolotinSuspended,DuSuspended,
ChenKdoping,JangIceScreening,HongPZT,MohiuddinHighk,ZhangSTM}

To date, most experimental studies have focused on probing the
carrier mobility $\mu$, or equivalently the transport scattering
time $\tau_t = m^*\mu/e$.\cite{TanMobility,SchedinGasMolecule,BolotinSuspended,DuSuspended,ChenKdoping,JangIceScreening,HongPZT,MohiuddinHighk}
Another important parameter in two-dimensional (2D) transport, the
quantum scattering time $\tau_q$, has not been well studied.
$\tau_q$ characterizes the momentum relaxation of a quasi-particle
and relates to its quantum level broadening $\Gamma$ through
$\Gamma=\hbar/2\tau_q$.

Quantitatively, the difference between $\tau_q$ and $\tau_t$ in
graphene is shown in the following equations:\cite{HwangTauq}
\begin{align}
\label{eq1}
&\frac{1}{\tau_q}=\int_0^\pi \mathrm{Q}(\theta)(1+\mathrm{cos}\theta)d\theta \\
&\frac{1}{\tau_t}=\int_0^\pi
\mathrm{Q}(\theta)(1+\mathrm{cos}\theta)(1-\mathrm{cos}\theta)d\theta.
\notag
\end{align}
Here, $\theta$ is the scattering angle and Q($\theta$) depends on
specific scattering mechanisms.\cite{DasSarmaTauq,HwangTauq} While
small-angle events weigh heavily towards $\tau_q$, $\tau_t$ is
mostly affected by right angle scatterings. Measurement of
$\tau_t/\tau_q$ has proven to be a powerful diagnostic tool in
revealing complex scattering scenarios in conventional 2D electron
gases (2DEGs).\cite{Harrang,DasSarmaTauq,ColeridgeSdHO,SyedGaN}
For example, short-ranged scattering sources give rise to
$\tau_t/\tau_q\sim1$ while charged impurities far away from a 2DEG
lead to predominately small-angle events, resulting in large
$\tau_t/\tau_q$. The former has been observed in silicon inversion layers and the latter characterizes modulation doped GaAs 2DEGs.\cite{Harrang,DasSarmaTauq,ColeridgeSdHO} Despite its demonstrated importance, the study of
$\tau_q$ in graphene has been scant. Existing data are largely
obtained from the linewidth of cyclotron resonance at low densities.\cite{JiangInfrared} A systematic comparison between $\tau_t$ and
$\tau_q$ has not been made.

In this work, we report a comprehensive study of $\tau_q$ in six
graphene samples over a wide range of carrier densities 1.2 $< n <
6\times10^{12}$/cm$^2$ and mobility 4,400 $< \mu <$ 17,000
cm$^2$/Vs. $\tau_q$ is obtained from  Shubnikov-de Haas (SdH)
oscillations and ranges approximately 25-74 fs in these samples,
corresponding to $\Gamma$ = 4.5-13 meV. The $n$-dependence of
$\tau_t$, $\tau_q$, and their ratio $\tau_t/\tau_q$ can all be
explained by a self-consistent Boltzmann transport theory\cite{AdamSelf,HwangTauq} using three parameters: the charged
impurity density $n_{imp}$, the impurity-graphene distance $z$, and
the resistivity from short-ranged scatterers $\rho_{short}$. Our
results indicate that the mobility in current graphene-on-SiO$_2$
samples is limited by scattering from charges residing within 2
nm of the graphene sheet. We speculate that charges present at the
graphene/SiO$_2$ interface are the major sources of scattering.

\begin{figure}[htpb]
\includegraphics[angle=0,width=2.5in]{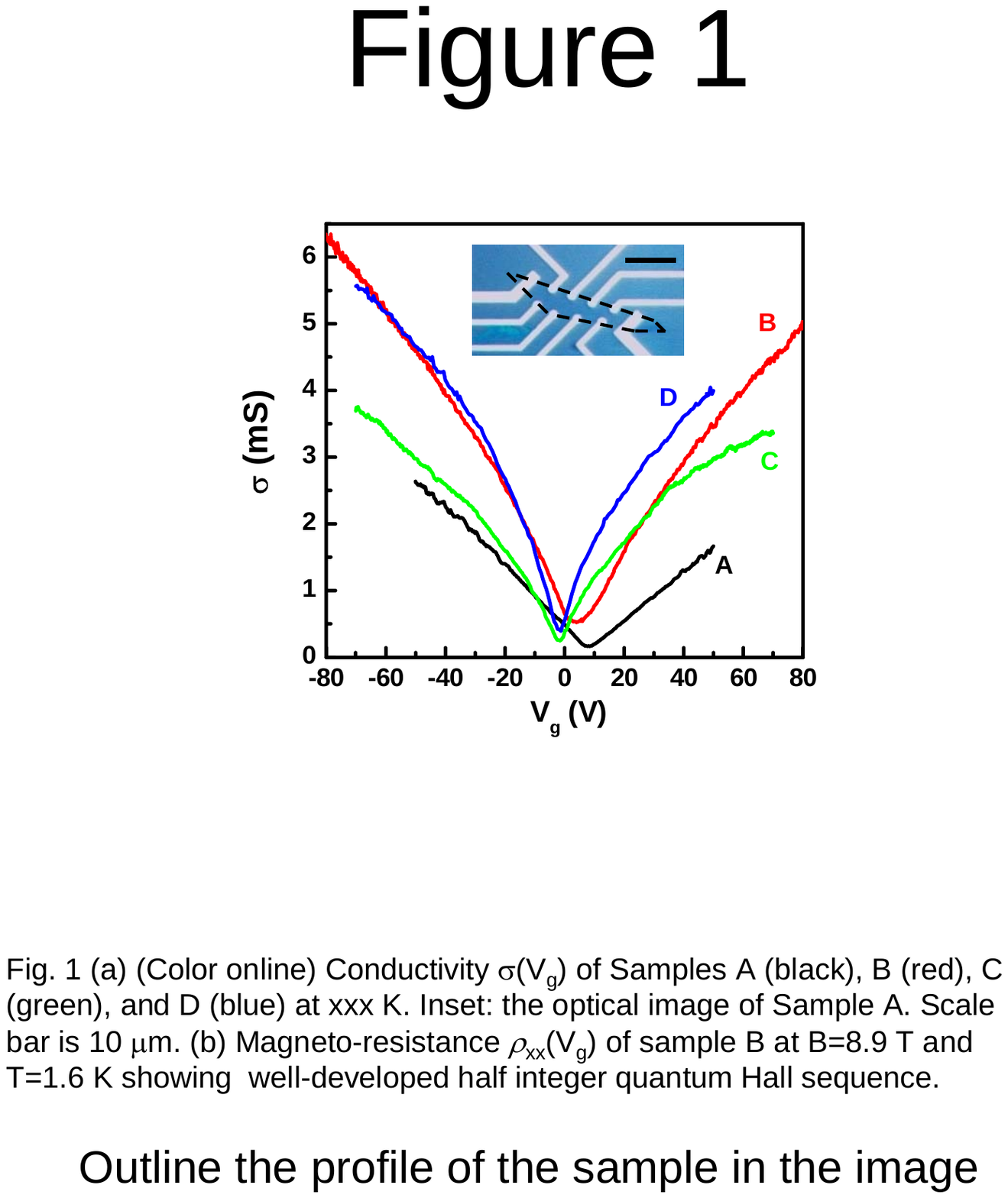}
\vspace{-0.1in} \caption[]{(Color online) $\sigma$(V$_\mathrm{g}$)
of samples A (black), B (red), C (green), and D (blue) below 10 K.
Inset: Optical image of sample A. Edge of the piece outlined. Scale
bar is 10 $\mu$m.\label{sample}}
\end{figure}

Single-layer graphene sheets are mechanically exfoliated onto 290 nm
SiO$_2$/doped Si substrates and identified optically. Rectangular
pieces are processed into Hall bar devices using standard e-beam
lithography followed by metal deposition (Fig.~\ref{sample} inset). The fabrication details are given in Ref.~\cite{supporting}.
Representative data from four samples (denoted as samples A-D) are
presented in details.

Transport experiments are performed in a pumped He$^4$ cryostat with
a base temperature of 1.4 K and equipped with a 9 T magnet. Standard
lock-in techniques are used with an excitation current of 50-200 nA.
The doped Si substrates serve as back gate electrodes, to which a
voltage (V$_\mathrm{g}$) is applied to tune the carrier density and
hence the conductance of graphene. We extract carrier density from
SdH oscillations and obtain a gating efficiency of
$\alpha$ = d$n$/dV$_\mathrm{g}$ = 7$\times$10$^{10}$/cm$^2$V of the
backgate. All measurements are taken at $T <$ 10 K to eliminate
electron-phonon scattering.

Figure~\ref{sample} shows the conductivity $\sigma$ vs
V$_\mathrm{g}$ taken on samples A-D. $\sigma$(V$_\mathrm{g}$) varies
linearly with V$_\mathrm{g}$ for the whole density range in sample A
and exhibits sublinear $\sigma$(V$_\mathrm{g}$) at high
V$_\mathrm{g}$ for samples B-D, where the "bowing" is most
pronounced in sample C. The qualitative features of these traces
resemble results reported previously,\cite{BolotinSuspended,DuSuspended,ChenKdoping,JangIceScreening} and have been
explained by a self-consistent Boltzmann theory combining scattering
from long and short-ranged sources:\cite{AndoGraphene,Nomura2007,AdamSelf}
\begin{align}
\label{eq2} &\frac{1}{\sigma}=\frac{1}{\sigma_{long}}+\rho_{short};
&\sigma_{long}=n e \mu_\mathrm{FE}+\sigma_{res}.
\end{align}
In this framework, $\rho_{short}$ denotes a constant contribution to
resistivity from short-ranged scattering sources such as defects or
neutral adsorbates. CI are thought to give rise to a linear
$\sigma$(V$_\mathrm{g}$), which implies a constant field effect
mobility $\mu_\mathrm{FE}$ and consequently $\tau_t \propto \sqrt{n}$.
Equations~\ref{eq2} produce excellent fittings to the
$\sigma(\mathrm{V_g})$ data of all our samples. The resulting
$\mu_\mathrm{FE}$ and $\rho_{short}$ span 4,400-17,000 cm$^2$/Vs and
40-165 $\Omega$ respectively (Table~\ref{tbl:list}), covering much
of the variations reported in the literature. The residue
conductivity $\sigma_{res}$ ranges 0.1-0.35 mS.\cite{ChenKdoping,TrushinSresidue} In samples exhibiting
electron-hole asymmetry, our analyses focus on the carrier type with
the higher $\mu_\mathrm{FE}$ to avoid complications associated with
contact doping.\cite{HuardContact}
\begin{table}
 \centering
 \begin{tabular} {|c|c|c|c|c|c|c|} \hline
 &$\mu_\mathrm{FE}$&$\rho_{short}$&$\tau_q$&$\tau_t/\tau_q$&$z$&$n_{imp}$\\
 &(cm$^2$/Vs)&($\Omega$)&(fs)& &(nm)&(10$^{11}$/cm$^2$)\\ \hline
A&4,400&40&31(38)&2.7(2.2)&0&10.4\\ \hline
B&10,000&55&33(49)&5.1(3.4)&2(1)&(7.7)\\ \hline
C&9,500&165&66&1.7&0&4.8\\ \hline D&17,000&105&53&3.5&2&7\\ \hline
 \end{tabular}
 \caption[]{$\mu_\mathrm{FE}$, $\rho_{short}$, $\tau_q$, $\tau_t/\tau_q$, $z$, and $n_{imp}$ for samples A-D. $\tau_q$ and $\tau_t/\tau_q$ are given for $n \sim$ 3 x 10$^{12}$/cm$^2$. The uncertainty in $z$ is 1-2 $\mathrm{\AA}$ for all samples. Data in parenthesis are after the corrections of density inhomogeneity.}
 \label{tbl:list}
\vspace{-0.1in}
\end{table}

Although Eqs.~\ref{eq2} provide a good description of existing
conductivity measurements, the origin of scattering sources in
graphene is still under debate. In addition to CI, ripples, resonant scatterers and midgap states are also potential candidates.\cite{KatsnelsonRipple,WiesendangerResonance,GuineaMidgap,KatsnelsonCluster} Unlike potassium adatoms,\cite{ChenKdoping} certain adsorbates seem to dope graphene but
not degrade its mobility.\cite{SchedinGasMolecule} The role of the
dielectric environment also appears controversial.\cite{JangIceScreening,MohiuddinHighk,HongPZT} While Jang $et$ $al$. find agreement with the model using ice as a top dielectric layer,\cite{JangIceScreening} Ponomarenko $et$ $al$. report screening effects much smaller than expected from the CI model using liquid dielectric layers.\cite{MohiuddinHighk} Within the
charged-impurity model, the origin of such impurities remain
unclear: Adsorbates on top of graphene, charges adsorbed/trapped at the graphene/SiO$_2$
interface or residing inside the substrate are all possible
candidates.

\begin{figure}[htpb]
\includegraphics[angle=0,width=3.2in]{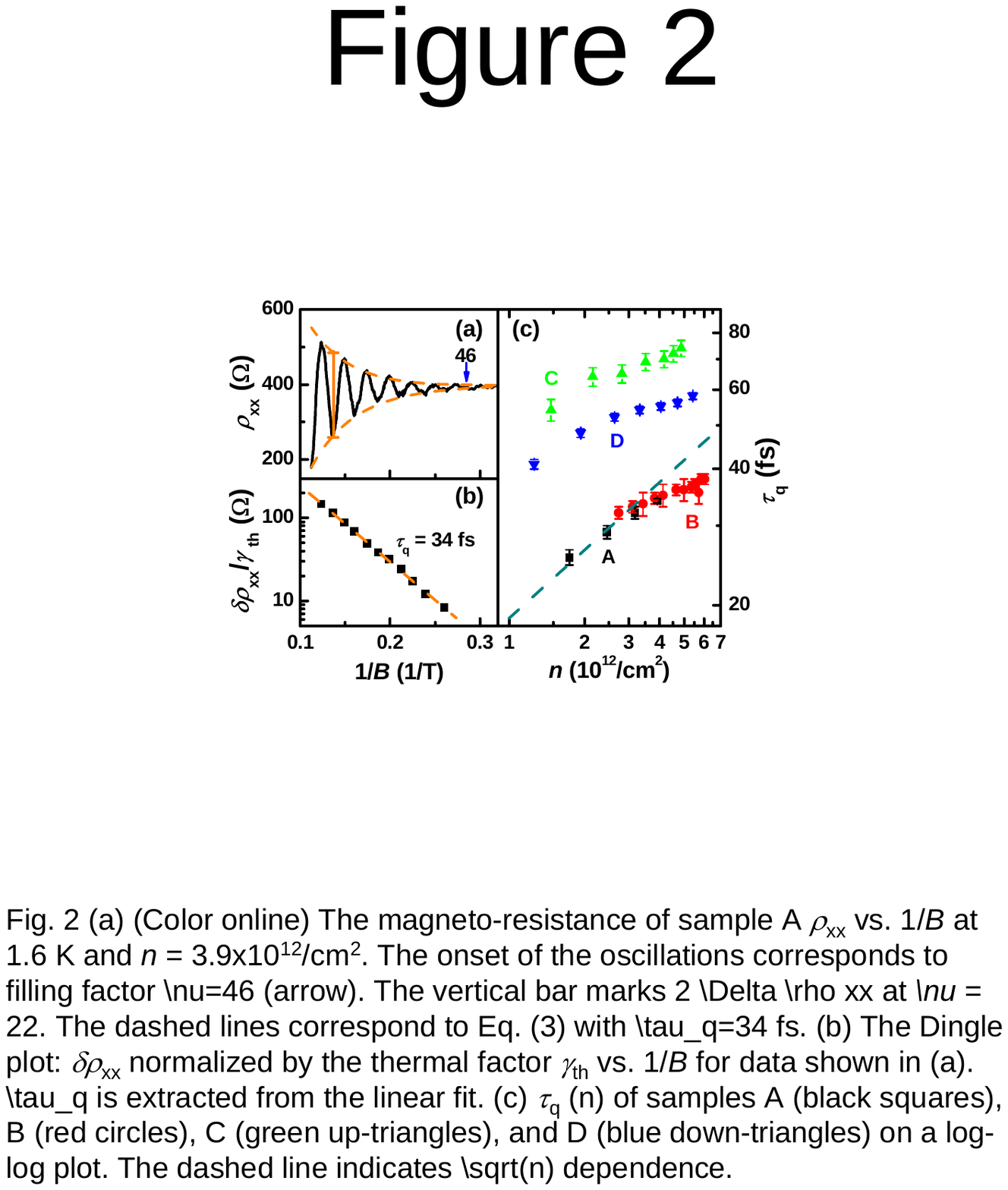}
\vspace{-0.1in}
 \caption[]{(Color online) (a) $\rho_\mathrm{xx}$(1/$B$) of sample A at 1.6 K and $n = 3.89\times10^{12}$/cm$^2$.
 The onset of the oscillations corresponds to filling factor $\nu$ = 46 (arrow).
 The vertical bar marks 2$\delta \rho_\mathrm{xx}$ at $\nu$ = 22.
 The dashed lines correspond to Eqs.~\ref{eq3} with $\tau_q$ = 34 fs.
 (b) The corresponding Dingle plot: $\delta \rho_\mathrm{xx}$/$\gamma_{th}$ vs. 1/$B$.
 $\tau_q$ is extracted from the linear fit. (c) $\tau_q$($n$) of samples A (squares), B (circles), C (up-triangles),
  and D (down-triangles) on a log-log plot. The dashed line indicates $\sqrt{n}$ dependence.
\label{tauq}}
\end{figure}

We have measured the quantum scattering time $\tau_q $ in graphene
to further address the above issues. We determine $\tau_q$ from the
magnetic field dependence of SdH oscillations following procedures
well-established in conventional 2DEGs.\cite{ColeridgeSdHO}
Figure~\ref{tauq}(a) shows the magneto-resistance
$\rho_\mathrm{xx}$($B$) of sample A at $n = 3.89\times
10^{12}$/cm$^2$ and $T$ = 1.6 K. The oscillatory amplitude $\delta
\rho_\mathrm{xx}$ can be described by:
\begin{equation}
\label{eq3}
\frac{\delta\rho_\mathrm{xx}}{\rho_0}=4\gamma_{th}\mathrm{exp}(-\frac{\pi}{\omega_c\tau_q});
\gamma_{th}=\frac{2\pi^2k_BT/\hbar\omega_c}{\mathrm{sinh}(2\pi^2k_BT/\hbar\omega_c)}.
\end{equation}
Here $\rho_0$ is the non-oscillatory background resistance,
$\gamma_{th}$ is the thermal factor, and $\omega_c$ the cyclotron
frequency in graphene. Here $m^* =E_F/v^2_F= \hbar\sqrt{\pi n}/v_F$
is the effective mass and $v_F= 1\times10^6$ m/s is the Fermi
velocity in graphene. Figure~\ref{tauq}(b) plots $\delta
\rho_\mathrm{xx}/\gamma_{th}$ vs. 1/$B$ in a semi-log plot (the
Dingle plot), where we extract $\tau_q$ = 34 fs from the slope of
the linear fit. The corresponding $\delta \rho_\mathrm{xx}$
calculated from Eqs.~\ref{eq3} is plotted in Fig.~\ref{tauq}(a) as
dashed lines and exhibits excellent agreement with data. In each
sample, the same procedure is repeated at different densities for $n
> 1.2\times10^{12}$/cm$^2$, where several well-developed SdH
oscillations are observed before the onset of quantum Hall states.
In some traces, a slowly varying background is subtracted before the
determination of $\delta \rho_\mathrm{xx}$, as described in
Ref.~\cite{supporting}.

Figure~\ref{tauq}(c) plots $\tau_q$($n$) in samples A-D (also listed
in Table~\ref{tbl:list} for $n = 3\times10^{12}$/cm$^2$).
$\tau_q$($n$) increases with increasing $n$ in all samples and spans
25-74 fs for 1.2$\times10^{12}$/cm$^2 < n < 6\times10^{12}$/cm$^2$.
These values correspond to a quantum level broadening $\Gamma$ =
4.5-13 meV and are in line with $\Gamma$ = 20-30 meV extracted from
the adsorption linewidth of cyclotron resonances at $n < 1\times10^{12}$/cm$^2$.\cite{JiangInfrared}

The $n$-dependence of $\tau_q$ in Fig.~\ref{tauq}(c) agrees
qualitatively with that of $\sigma$(V$_\mathrm{g}$) in
Fig.~\ref{sample}. This can be seen by separating the long and
short-ranged components in $\tau_{t,q}$ using the following
equations:\cite{HwangTauq}
\begin{align}
\label{eq4}
&\frac{1}{\tau_{t,q}} = \frac{1}{\tau_{t,q}^{long}}+\frac{1}{\tau_{t,q}^{short}};\\
&\tau_t^{short} = \frac{m^*}{n e^2 \rho_{short}};\enspace \frac{\tau_t^{short}}{\tau_q^{short}} = 1.1.\notag
\end{align}
Due to a small $\rho_{short}$, both $\tau_t$ and $\tau_q$ in sample
A follow closely the $\sqrt{n}$ dependence expected for charged
impurities residing in the graphene plane ($z$ = 0).\cite{HwangTauq}
Large $\rho_{short}$ and higher $\mu_\mathrm{FE}$ in samples B-D
(Table~\ref{tbl:list}) cause both scattering times to deviate from
the $\sqrt{n}$ dependence. However, $\tau_q$ does not correlate with
$\mu_\mathrm{FE}$ in a simple relation. Samples B and C exhibit similar
$\mu_\mathrm{FE}$, but their $\tau_q$ differ by a factor of 2.
Samples A and B show comparable $\tau_q$s despite the significant
difference in $\mu_\mathrm{FE}$. Sample C, with a moderate
$\mu_\mathrm{FE}$ and the highest $\rho_{short}$, exhibits the
highest $\tau_q$.

\begin{figure}[htpb]
\includegraphics[angle=0,width=3.2in]{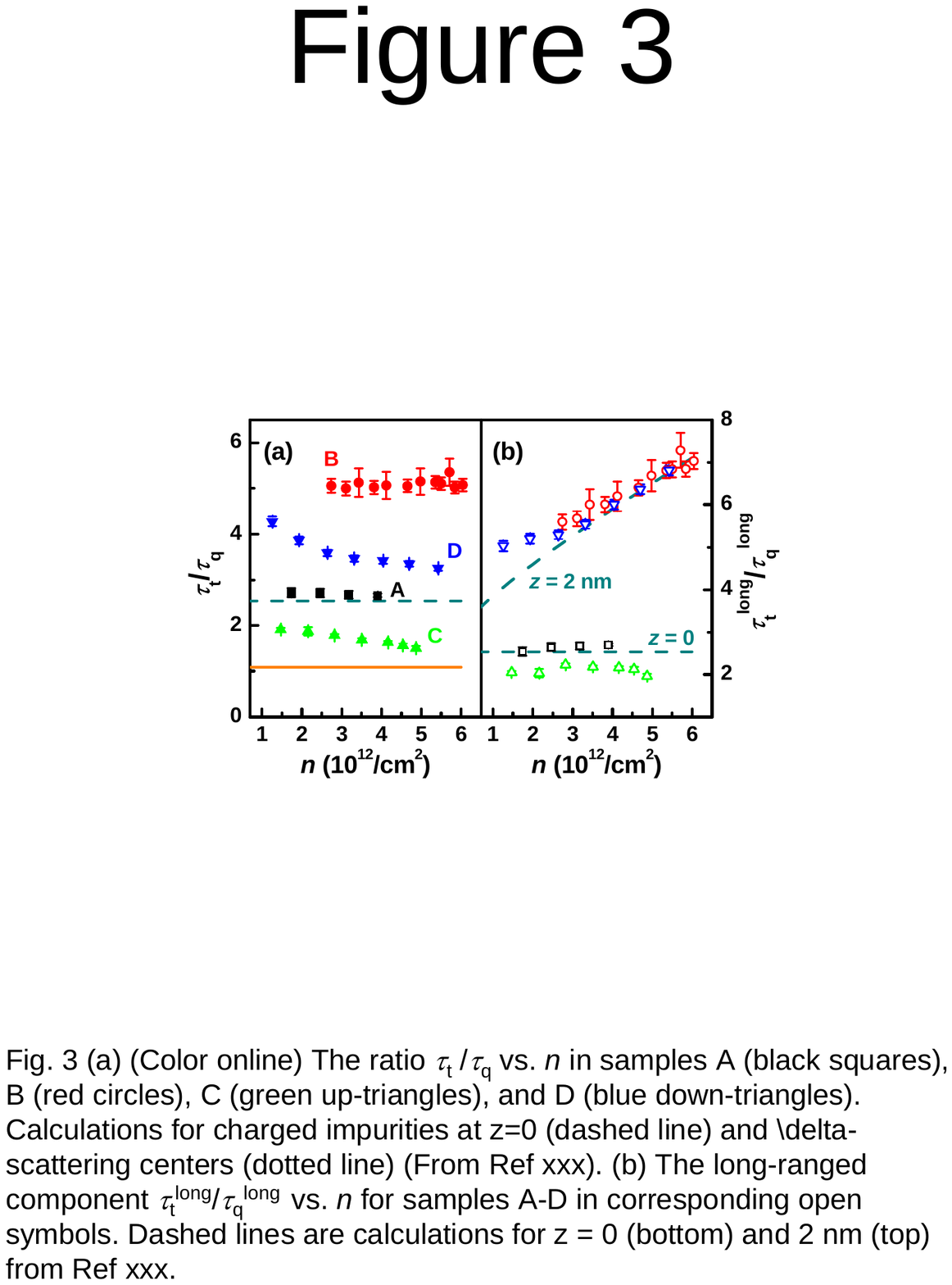}
\vspace{-0.1in} \caption[]{(Color online) (a) $\tau_t /\tau_q$ vs.
$n$ in samples A (squares), B (circles), C (up-triangles), and D
(down-triangles) along with calculations for CI at $z$ = 0 (dashed
line) and $\delta$-scattering centers (solid line)
(\cite{HwangTauq}). (b) Their long-ranged component
$\tau_t^{long}/\tau_q^{long}$  vs. $n$ in corresponding open
symbols. Dashed lines are calculations for $z$ = 0 (bottom) and 2 nm
(top) (\cite{HwangTauq}).  \label{ratio}}
\vspace{-0.28in}
\end{figure}

To understand the above observations, we calculate the ratio
$\tau_t/\tau_q (n)$ and the long-ranged component $\tau_t^{long}/\tau_q^{long}(n)$ using
Eqs.~\ref{eq4}, and plot the results in Figs.~\ref{ratio} (a) and
(b) (also Table~\ref{tbl:list}). Samples A and B show $n$-independent
$\tau_t/\tau_q$ of 2.7 and 5.1 respectively. For the other two samples, $\tau_t/\tau_q$
decreases slightly with increasing $n$, varying from 1.9 to 1.5 in
sample C and 4.3 to 3.3 in sample D. Clearly, the angular distribution Q($\theta$) in Eqs.~\ref{eq1} differs significantly in these samples.

By evaluating $\tau_t^{long}/\tau_q^{long}(n)$, we find that such variation may be naturally explained by varying the impurity-graphene distance $z$ within the
charged-impurity model. Theoretical calculations of short and long-ranged ratios are given in
solid and dashed lines, respectively, in Figs.~\ref{ratio} (a) and
(b).\cite{AdamSelf,HwangTauq} According to this model, the
$\tau_t^{long}/\tau_q^{long}(n)$ of sample A provides further
evidence for the domination of CI located in the graphene plane ($z$
= 0), where a constant 2.5 is expected (dashed
line).\cite{AdamSelf,HwangTauq} The $\tau_t/\tau_q (n)$ in sample C falls between the dashed and solid lines, which is the result of a large short-ranged component. Its $\tau_t^{long}/\tau_q^{long}(n)$ exhibits an approximate constant ratio of 2.1, also pointing to CI located in the graphene plane
(Fig.~\ref{ratio}(b)). $\tau_t^{long}/\tau_q^{long}$ in samples B
and D range from 5-7 and are best described by CI located 2 nm away
from the graphene sheet (Fig.~\ref{ratio}(b)). In all six samples, we
find 2 $< \tau_t^{long}/\tau_q^{long} <$ 7, corresponding to CI
located within 2 nm of the graphene sheet.

The knowledge of the impurity-graphene distance $z$ is essential in
correctly determining the CI density $n_{imp}$ in the vicinity of
graphene. In the CI model, $n_{imp}$ is related to $\mu_\mathrm{FE}$
through $\mu_\mathrm{FE}$= $C/n_{imp}$, where $C$ =
5$\times10^{15}$/Vs is $n$-independent for $z$ = 0.\cite{ChenKdoping} At a finite $z$, $C$ increases with increasing
$n$ due to screening.
Using the equations in Refs.~\cite{AdamSelf,HwangTauq}, we
numerically calculate $C$($n$, $z$) and estimate $n_{imp}$ by
fitting $\sigma(n)$ to Eq.~\ref{eq2}. The results are listed in
Table~\ref{tbl:list} while an exemplary fitting is given in
Ref.~\cite{supporting}. Clearly, $\mu_\mathrm{FE}$ (or $\tau_t$) is
affected by both $z$ and $n_{imp}$. For example, the difference in
$\mu_\mathrm{FE}$ between samples A and D mainly stems from $z$,
instead of $n_{imp}$. In contrast, $\tau_q$ serves as a better
measure of $n_{imp}$ due to its weaker dependence on $z$.

Next we briefly assess the effect of density inhomogeneity $\delta
n$ on the measurement of $\tau_q$. Caused by CI, $\delta n$ measures approximately a
few 10$^{11}$/cm$^2$ near the Dirac point,\cite{MartinSTMpuddle,ZhangSTM} and is expected to decrease with increasing $n$ due to electron screening.\cite{RossiInhomogeneity}
$\delta n$ introduces phase smearing in $\rho_\mathrm{xx}(B)$, effectively
reducing the SdH oscillation amplitude and suppressing the value of
$\tau_q$ determined through the Dingle plot, as demonstrated in GaN
2DEGs.\cite{SyedGaN} We have used the intercept of the Dingle plot
at 1/$B$ = 0 as a criterion\cite{ColeridgeSdHO} to obtain $\delta
n$ and the corresponding corrections to $\tau_q$. Details are given
in Ref.~\cite{supporting}. We estimate $\delta n$ to be
$\sim7\times10^{10}$/cm$^2$ in sample A, and $\sim
9\times10^{10}$/cm$^2$ in sample B. These estimates are consistent
with the highest filling factors observed in these samples ($\nu$ =
46 for sample A and $\nu=74$ for sample B). Overall, $\delta n$/$n$ decreases
rapidly with $n$, in agreement with theory; but the magnitude is only a few percent in the density range we studied, which is significantly smaller than the theoretical predictions.\cite{supporting,RossiInhomogeneity,PoliniInhomogeneity} The above correction leads to
$\sim20\%$ increase of $\tau_q$ in sample A and 50$\%$ in sample B.
The corrected $\tau_q$s are given in Table~\ref{tbl:list} in
parenthesis. In samples C and D, the corrections are smaller than
the error bars of $\tau_q$ and therefore omitted.
$\tau_t^{long}/\tau_q^{long}$ in sample B now corresponds to CI
located at $z$ = 1 nm instead of previously determined $z$ = 2 nm,
but the main picture does not change.\cite{supporting}

Our study of $\tau_q$ and $\tau_t/\tau_q$ provides critical
information in differentiating various scattering scenarios in
graphene.\cite{HwangTauq,KatsnelsonRipple,WiesendangerResonance,GuineaMidgap,KatsnelsonCluster} A detailed comparison
to theory is only made for the CI model at this point, but can be
extended to other proposals as quantitative predictions become
available. The diverse behavior our samples exhibit can all be
understood very well within the CI model using three parameters:
$n_{imp}$, $z$, and $\rho_{short}$. We speculate that uncontrolled
spatial variation of SiO$_2$ surface properties, as well as sample
preparation conditions (e. g. humidity) may have been the primary
reasons behind the observed differences among samples, although
variations in preparation procedures cannot be ruled
out.\cite{supporting}

Our results indicate that the dominant CI reside within 2 nm of, and
sometimes in the immediate vicinity of the graphene sheet. Primary
candidates of this nature are charges carried by adsorbates on top
of graphene and/or at the graphene/SiO$_2$ interface. The role of
adsorbates is especially highlighted in the current annealing
treatment of suspended graphene.\cite{BolotinSuspended,DuSuspended}
In the literature, various approaches, including resist-free
processing,\cite{ZhangSTM,StaleyDry,GiritDry} UHV baking,\cite{JangIceScreening,ZhangSTM} and current annealing,\cite{MoserCurrent} have been used to remove contaminants on top
of graphene without significant improvement to mobility. These
observations collectively suggest that adsorbates on top of graphene
cannot be the major culprit in limiting mobility at the current
level. Instead, we speculate that charges (e. g. Na$^+$) and
molecular groups (e. g. OH) adsorbed/trapped on the SiO$_2$ surface
prior to the exfoliation of graphene are the major source of
scattering. The $z$ = 0 found in samples A and C lends strong
support to this hypothesis. Moreover, the small $z$ observed in
other samples can be accounted for by the existence of a spacer
layer between graphene and SiO$_2$ (e.g. H$_2$O). The evidence of
such a layer is widely observed in AFM height measurements of
graphene. In this scenario, the concentration of adsorbed charges
($n_{imp}$), together with the thickness of the spacer layer ($z$)
can account for the wide span of mobility seen in our samples. It
may also explain why graphene on a variety of substrates displays a
similar range of mobility\cite{MohiuddinHighk} since the bulk
properties of these substrates are less relevant here.

Finally we note that the above determined $z$ can be expanded to
represent an average impurity-graphene distance. Using this concept,
we consider the contribution of uniformly distributed charges within
the bulk of the SiO$_2$ substrate. Our simulations show that oxide
charges in commercially available SiO$_2$ are unlikely to be a major
source of scattering at the current mobility
level.\cite{supporting}


In conclusion, we have systematically studied the quantum and
transport scattering times in graphene. Our data will prove useful
in critical examinations of existing scattering scenarios. Within
the CI model, the ratio of $\tau_t/\tau_q$ indicates that charged
impurities residing within 2 nm of the graphene sheet are the main
sources of scattering in graphene. Such information provides
important guidance to the effort of improving carrier mobility in
graphene.

\begin{acknowledgments}
We are grateful for helpful discussions with S. Adam and L. Song,
and technical assistance from S.-H. Cheng and S. Syed. We thank P.
Eklund for providing access to his Raman spectrometer. Work at Penn
State is supported by NSF Grants No. ECS-0609243, No. CAREER
DMR-0748604, and No. MRSEC DMR-0820404. The authors acknowledge use of facilities at the PSU
site of NSF NNIN.
\end{acknowledgments}

\bibliography{scatteringtime}

\end{document}